# Drug transport mechanism of P-glycoprotein monitored by single molecule fluorescence resonance energy transfer

S. Ernst[a§], B. Verhalen[b§], N. Zarrabi[a], S. Wilkens*[b], M. Börsch*[a]

[a] 3rd Institute of Physics, University of Stuttgart, Pfaffenwaldring 57, 70550 Stuttgart, Germany.
[b] Department of Biochemistry & Molecular Biology, SUNY Upstate Medical University,
750 East Adams Street, Syracuse, NY 13210, USA.

**ABSTRACT**

In this work we monitor the catalytic mechanism of P-glycoprotein (Pgp) using single-molecule fluorescence resonance energy transfer (FRET). Pgp, a member of the ATP binding cassette family of transport proteins, is found in the plasma membrane of animal cells where it is involved in the ATP hydrolysis driven export of hydrophobic molecules. When expressed in the plasma membrane of cancer cells, the transport activity of Pgp can lead to the failure of chemotherapy by excluding the mostly hydrophobic drugs from the interior of the cell. Despite ongoing effort, the catalytic mechanism by which Pgp couples MgATP binding and hydrolysis to translocation of drug molecules across the lipid bilayer is poorly understood. Using site directed mutagenesis, we have introduced cysteine residues for fluorescence labeling into different regions of the nucleotide binding domains (NBDs) of Pgp. Double-labeled single Pgp molecules showed fluctuating FRET efficiencies during drug stimulated ATP hydrolysis suggesting that the NBDs undergo significant movements during catalysis. Duty cycle-optimized alternating laser excitation (DCO-ALEX) is applied to minimize FRET artifacts and to select the appropriate molecules. The data show that Pgp is a highly dynamic enzyme that appears to fluctuate between at least two major conformations during steady state turnover.

## 1 INTRODUCTION

According to the World Health Organization, cancer is the second most common cause of death (after cardiovascular disease) in the majority of developed countries, resulting in about 8 million deaths worldwide each year (http://www.who.int/cancer/en/). A common approach to battle cancer is chemotherapy. While chemotherapy is a very powerful weapon against metastatic tumors, its success is often limited by a property of the cancer known as multidrug resistance (MDR). MDR means that the cancerous cells are resistant or become resistant to a wide spectrum of cytotoxic drugs used in chemotherapy. One of the main reasons for MDR is the expression of energy dependent transport proteins in the plasma membrane of tumor cells which prevent the entry of the cytotoxic chemotherapeutic drugs into the cytosol. The first such drug transporter described was P-glycoprotein (Pgp). Pgp is naturally expressed at high levels in kidney, liver, placenta and in cells responsible for the blood-brain barrier where its natural substrates are neutral or cationic organic molecules including environmental toxins, by-products of the everyday metabolism of foodstuffs[1].

...................................................................................................................................................................

[§] Equal contribution

[*] wilkenss@upstate.edu; phone (1) 315 464 8703; fax (1) 315 464 8750; http://www.upstate.edu/biochem/faculty.php?ID=wilkenss
  m.boersch@physik.uni-stuttgart.de; phone (49) 711 6856 4632; fax (49) 711 6856 5281; http://www.m-boersch.org

Pgp belongs to the superfamily of the adenosine triphosphate (ATP) binding cassette (ABC) transporters. This class of transporters mediate active translocation across the cell membrane and can be found in all prokaryotic and eukaryotic cells. The transporters couple the hydrolysis of ATP to translocation of a broad range of substrates across the membrane[2]. ABC transporters are characteristically composed of two transmembrane domains (TMDs) and two nucleotide binding domains (NBDs). In Pgp, the domain organization given by the primary sequence is N-TMD1-NBD1-TMD2-NBD2-C. The TMDs of Pgp are each composed of six transmembrane α helices. ATP hydrolysis takes place at the interface of the two NBDs while the transport channel in which the substrate crosses the lipid bilayer is formed by the two TMDs. Together, the four domains form a transporter with pseudo two-fold symmetry that extends 136 Å perpendicular and 70 Å parallel to the plane of the membrane[3] (Fig. 1).

Due to the important role of Pgp in cancer research, Pgp is one of the best characterized ABC transporters. Based on a wealth of structural data from electron microscopy[4, 5] and X-ray crystallography[6], together with results from biochemical and biophysical experiments including chemical cross-linking, fluorescence and EPR spectroscopy as well as nucleotide, drug and inhibitor binding studies, a picture of the catalytic mechanism of Pgp is emerging in which binding and hydrolysis of MgATP in alternating catalytic sites on the NBDs is coupled to drug binding and transport via out- and inward facing conformations of the two transmembrane domains[7-12]. A schematic of the in- and outward facing conformations is shown in Figure 1A.

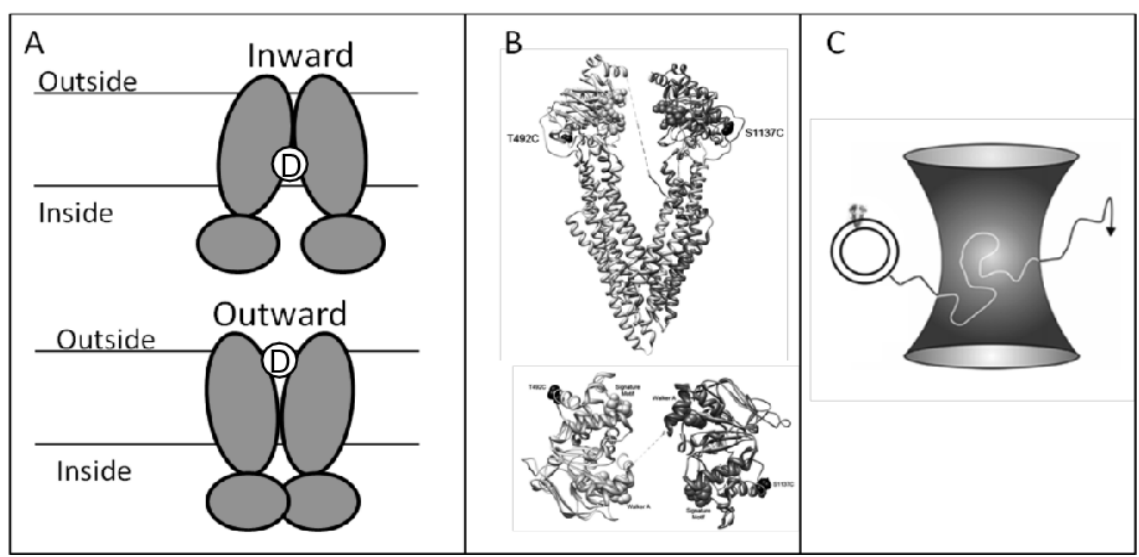

**Figure 1: Basic mechanism of P-glycoprotein (Pgp) and experimental design of single-molecule fluorescence resonance energy transfer (FRET) to monitor conformational changes in solution.** (**A**) Information gained from structural studies show two main catalytic states of ABC transporters, the outward and inward facing conformation. The inward conformation is where the TMDs are open towards the cytosol. For exporters such as Pgp, this is thought to allow the transporter to acquire substrate (drug) from the inner leaflet of the bilayer and then move the substrate towards the outside of the cell, depicted by the outward facing conformation. (**B**) The Pgp structure[3] (Aller, 2009, PDB ID 3g5u) is displayed with the Walker A motif and Signature Sequence in space fill (notated in the top view of Pgp, bottom panel). Also in space fill are the two residues genetically modified to cysteines for fluorophore labeling (Thr 492 and Ser 1137). (**C**) A proteoliposome containig a single Pgp is freely diffusing through the confocal volume of the lasers drawn as gray cone.

Details of this mechanism, such as the order of the various steps and the degree of motion of the NBDs, however, are still poorly understood. To delineate the conformational changes Pgp is undergoing during the transport cycle, we developed a single-molecule fluorescence approach (as previously reported to observe the rotary subunit movements in $F_oF_1$-ATP synthase[13-19]) that allows monitoring of NBD dynamics in individual membrane bound transporters. In this assay, proteoliposomes (with a diameter of 70-150 nm) that contain a single Pgp labeled with donor and acceptor dyes (Fig. 1B) are allowed to diffuse freely through a confocal volume of focused laser light (Fig. 1C). We apply single-molecule Förster-type fluoresence resonance energy transfer (FRET) to monitor the distance changes between the two nucleotide binding domains. The well-known distance dependence of FRET is widely used to map the spacing between two intramolecular or intermolecular positions in the range between 2 and 8 nm[17, 20-25]. Based on recent X-ray crystal structures of ABC transporters in general and Pgp in particular, conformational changes during the transport cycle are predicted to result in a sequence of distance changes between the two NBD bound fluorophores, leading to a step wise change in FRET efficiency. Real time monitoring of FRET efficiencies with millisecond time resolution and sub-nanometer precision is a powerful tool for gaining a deeper understanding of the structural changes that couple binding and hydrolysis of MgATP to binding and translocation of drug molecules. In this work, we conducted single-molecule FRET experiments using Pgp double mutant TS, in which Thr 492 in NBD1 and Ser 1137 in NBD2 were replaced by cysteine residues for labeling with maleimide linked fluorescent dyes (Figure 1B; dark grey). In a series of experiments, we monitored FRET efficiencies of single Pgp molecules under conditions of 'apo' (no drugs or nucleotides), 'transport' (MgATP + verapamil), 'vanadate trapped' (MgATP, verapamil + orthovanadate) and modulation by cyclosporin A. The data show that Pgp fluctuates between two major conformations, one with high and one with lower FRET efficiency during drug stimulated ATP hydrolysis. Additional states are observed that are less populated, possibly because they are short lived. The findings are discussed in context of the available structural and biochemical data and a model of the transport mechanism of Pgp is presented.

## 2 EXPERIMENTAL PROCEDURES

### 2.1 Sample preparation

Single-molecule FRET experiments were carried out with P-glycoprotein labeled on NBD1 and NBD2 with donor and acceptor dyes. Mouse mdr3 (Pgp) was expressed in *Pichia pastoris* and purified as described by Lerner-Marmarosh et al.[26], with minor modifications. After anion exchange with DE-52 resin, protein was concentrated and applied to a Superdex 200 size exclusion chromatography column (10/30; GE Healthcare) in 10 mM Mops, 50 mM NaCl, 10% glycerol, 0.5 mM EDTA, 0.5% CHAPS, 0.1 mM TCEP, pH 7.0. The protein was mixed with a molar excess of Alexa488-maleimide and Atto610-maleimide and incubated for one hour at 20 °C. Binding of excess dye was terminated by addition of 1 mM cysteine for 10 minutes at 20 °C. Labeling efficiency was determined by UV-Vis absorption spectroscopy. The protein to dye ratio needed for stoichiometric labeling was determined by testing multiple ratios. The labeled protein was then added to a 19:1 PC:PA lipid mix in the ratio of 1:1739 protein:lipid (w/w). The protein was reconstituted into proteoliposomes *via* gel filtration using a Sephadex G50 gravity flow column.

### 2.2 Single-molecule FRET measurements

For the excitation of the two FRET fluorophores two lasers were used. For the FRET donor excitation a fiber coupled picosecond laser at 488 nm (PicoTa 490, up to 80 MHz repetition rate, Picoquant) was applied. The second HeNe laser provided continuous-wave laser light at 594 nm (Coherent, 05-LYR-173). The laser was switched in the nanosecond range *via* an acousto-optical modulator (AOM, model 3350-192, Crystal technologies). Both lasers were triggered and synchronized by an programmable external pulse generator[27]. The intensity for the FRET donor excitation laser at 488 nm was attenuated to 150 µW, and the laser power for the acceptor test at 594 nm was set to 30 µW to avoid photobleaching of the fluorophores.

In order to obtain an enlarged confocal volume for the solution measurements, the two laser beams were compressed by a pair of lenses. For the 488 nm laser beam, a set of semi concave lenses with a focus of 250 mm and 40 mm were used. The 594 nm laser beam was compressed by a factor of 4:1. Both were overlaid manually using a dichroic beam splitter (DCXR 488, AHF Tübingen). The correct position of the two focal spots was controlled by scanning multicolor fluorescent beads with a diameter of 1 µm (transfluospheres 488/635, Molecular Probes) with both excitation lasers. The scanning procedure was accomplished by an x-y piezo scanner plus piezo objective positioner (Physik Instruments)[28, 29]. The digital controller of the scanner was addressed by a custom made LabView software. The center of the Gaussian intensity distributions was determined and the distance of the foci was minimized. The automatic analysis was performed by custom scripts written in MATLAB. The microscope setup is drawn in Fig. 2.

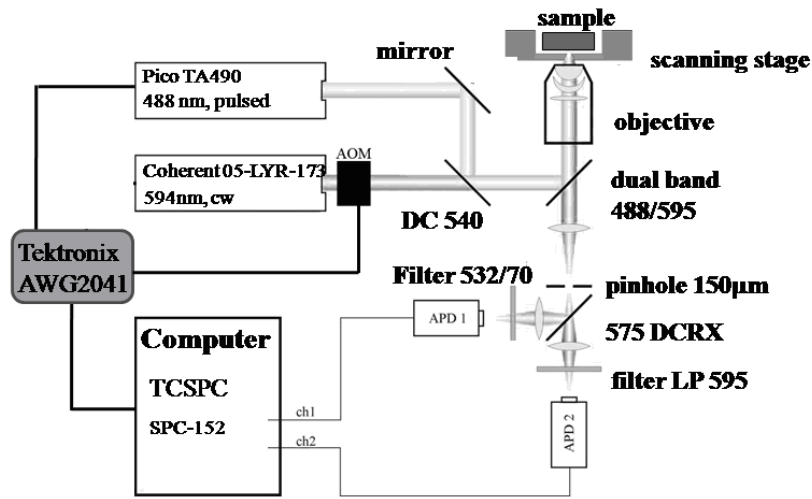

**Figure 2: Single-molecule FRET microscopy setup.** For the FRET experiments, a pulsed laser at 488 nm (FRET donor excitation) and an AOM-switched He/Ne laser at 594 nm for the FRET acceptor test was used.

The measurement setup is based on a custom built inverted confocal microscope (Olympus IX71). Laser excitation and fluorescence detection pathways were modified compared to previously reported detection schemes[30-38]. The laser beams were redirected by a dichroic filter (488/595, AHF) after entering the epifluorescent port of the microscope and focused into the solution by a water immersion objective (UPlanSApo 60xW, 1.2 N.A., Olympus). The fluorescent signal of the sample passed a 150 μm pinhole and was separated into two different spectral ranges with a dichroic beam splitter (DCXR 575, AHF). The emerging photons of both spectral ranges were recorded simultaneously by two avalanche photo diodes (AQR-14, Perkin Elmer). The fluorescent signal of Alexa488 was detected between 497 and 567 nm (HQ 532/70, AHF) and the signal of Atto610 from 595 nm (LP595, AHF). Single photons were registered by two synchronized TCSPC cards (SPC-152, Becker & Hickl) and simultaneously by a multi-channel counter card for imaging (NI.PCI 6602, National Instruments). With this setup it was possible to use the microtime tag of each photon registered with picosecond resolution, that is, related to the synchronization pulse of the TCSPC electronics, to obtain the FRET efficiency[29].

The time trajectories of the fluorescence signals were further analyzed with our custom software "Burst-analyzer"[36]. Single-molecule measurements with solutions of donor and acceptor dye only served for background corrections, and the fluorescence intensities of both detecting channels were adapted accordingly. Briefly, for FRET donor and acceptor channels, a background of 11 and 5 counts/ms were subtracted, respectively. Also the detection efficiencies of both detection channels as well as the quantum yields of the two dyes were taken into account. The correction factor $\gamma$ of the single-molecule FRET experiments was determined to be 0.79.

In order to select only single Pgp molecules with the correct combination of FRET fluorophores linked to the nucleotide binding domains (one donor and one acceptor), duty cycle-optimized alternating laser excitation (DCO-ALEX) was used[39] (Fig. 3A). With this approach we could eliminate photon bursts originating from incorrectly labeled Pgp molecules that otherwise could have been misinterpreted as apparent FRET signals, that is, Ppg labeled with donor only, acceptor only, or doubly labeled with either two donors or two acceptors. By gating the microtime, histograms of both TCSPC channels were separated (Fig. 3B). The time trajectories for the FRET traces and the acceptor test traces were reconstructed for each fluorophore based on the laser pulses.

For the correct sorting of photons (i.e. depending on the exciting laser pulse) a fast triggering of the two lasers and the photon counting electronics in the nanosecond range is required. This was achieved by an arbitrary waveform generator (AWG2041, Tektronix) with a time resolution of 1 ns. With this device we were able to optimize the duty cycle for each laser independently and additionally trigger both TCSPC cards.

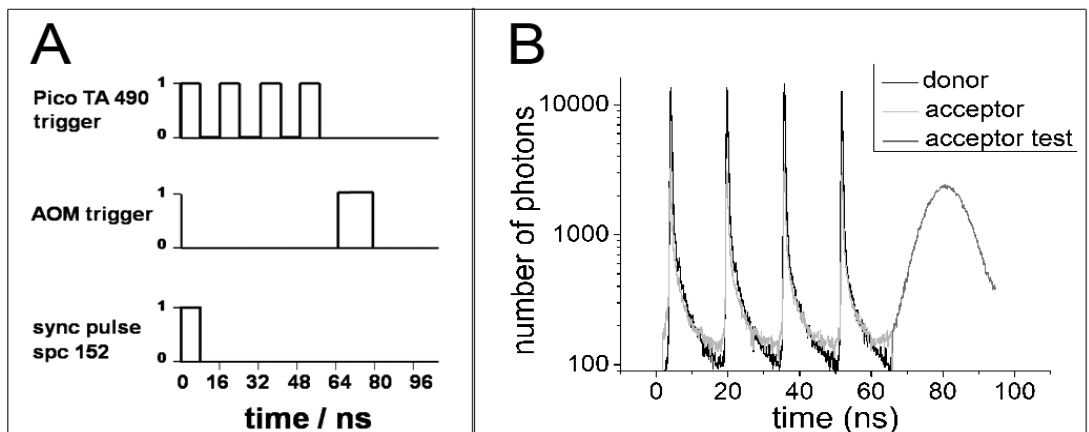

**Figure 3: TTL pulse sequences generated by the arbitrary waveform generator for duty cycle-optimized alternating pulsed laser excitation.** (**A**) Trigger pulse sequence for the PicoTA 490, the AOM for the HeNe laser, and the synchronization of the TCSPC cards. (**B**) Microtime histogram of the photons as detected by the two APDs in one period of 96 ns. The FRET measurement was completed in the first 64 ns of the 96 ns cycle. In the second time window (from 64 to 96 ns) a second laser pulse was applied to probe the acceptor dye.

The single-molecule FRET experiment was divided into two time windows. In the first time window, the FRET measurement was carried out with four consecutive pulses of the exciting laser for the FRET donor. Each laser pulse was 60 ps long and delayed by 16 ns. In the second time window, one laser pulse with the 594 nm wavelength was used in order to install a so called "acceptor test" for a duration of 16 ps (Fig. 3). This kind of test allows incorrectly labeled proteins (singly labeled or doubly labeled with either the donor or acceptor only) to be filtered out. The interval of the second time window was 32 ns. A 16 ns pulse for the AOM switched the HeNe laser beam on, and the following 16 ns pause switched it off. With this duty cycle scheme the complete fluorescence decay of both dyes was recorded. The maximum excitation rate of the FRET donor of 83.3 MHZ dropped down to 51% (i.e. 42.5 MHz) instead of down to only 26% (21.7 MHz) in a standard alternating laser excitation approach.

## 3 RESULTS

To monitor the transport cycle of a single Pgp molecule using single-molecule FRET, we attached donor and acceptor fluorophores to the protein, one to each NBD. For maleimide based labeling, two cysteines were introduced at positions T492 and S1137 (mutant 'TS') by site directed mutagenesis. After the labeling reaction Pgp was reconstituted into PC:PA (19:1) liposomes with a diameter of 70 to 150 nm. Two alternating lasers at 488 nm and 594 nm were used to excite the FRET donor and to probe the presence of the FRET acceptor, respectively. The laser pulse sequence was optimized for a high duty cycle to excite the FRET donor. Briefly, a series of four pulses at 488 nm was followed by one AOM-switched pulse at 594 nm.

Fig. 4 shows the photon bursts of a single Pgp transporter. In the lower panels, the fluorescence intensity trajectories of FRET donor Alexa488 and of FRET acceptor Atto610 on both NBDs are shown. The black traces show the donor intensity, the light gray the FRET signal and the dark gray trace displays the corresponding FRET acceptor intensity trajectory upon direct excitation, confirming the existence of the Atto610 dye. In the upper panels the distance of both fluorophores is displayed. FRET distances in each time interval were calculated by the following formula:

$$E_{FRET} = \frac{I_A}{\gamma I_D + I_A} = \frac{R_0^6}{R_0^6 + R_{DA}}$$

where $I_A$ and $I_D$ are the background-corrected fluorescence intensities in the FRET acceptor and FRET donor channels and $\gamma$ the correction factor combining quantum yields of the dyes and the detection efficiencies of both detection channels. $R_0$ is the Förster distance and $R_{DA}$ the distance between donor and acceptor fluorophores. For the FRET pair used here (Alexa488 and Atto610), the Förster radius was calculated to $R_0 = 5.7$ nm with a $\gamma$ factor of 0.79.

For the Pgp in Fig. 4A no nucleotide was added before the measurement ('Apo'). A mean FRET distance of 6.1 nm only displays minor fluctuations in the absence of substrates while the liposome was in the confocal focus between 50 and 75 ms as shown in the FRET distance trace above. However, in the presence of 200 μM verapamil and 1 mM MgATP ('Transport') the fluorescence intensities and the FRET efficiency changed (Fig. 4B). In the first 21 ms after the liposome entered the confocal focus the FRET distance was around 5.5 nm. After that the distance of the two NBDs slightly increased, followed by a sharp increase of FRET efficiency (or decrease in distance) to around 4 nm.

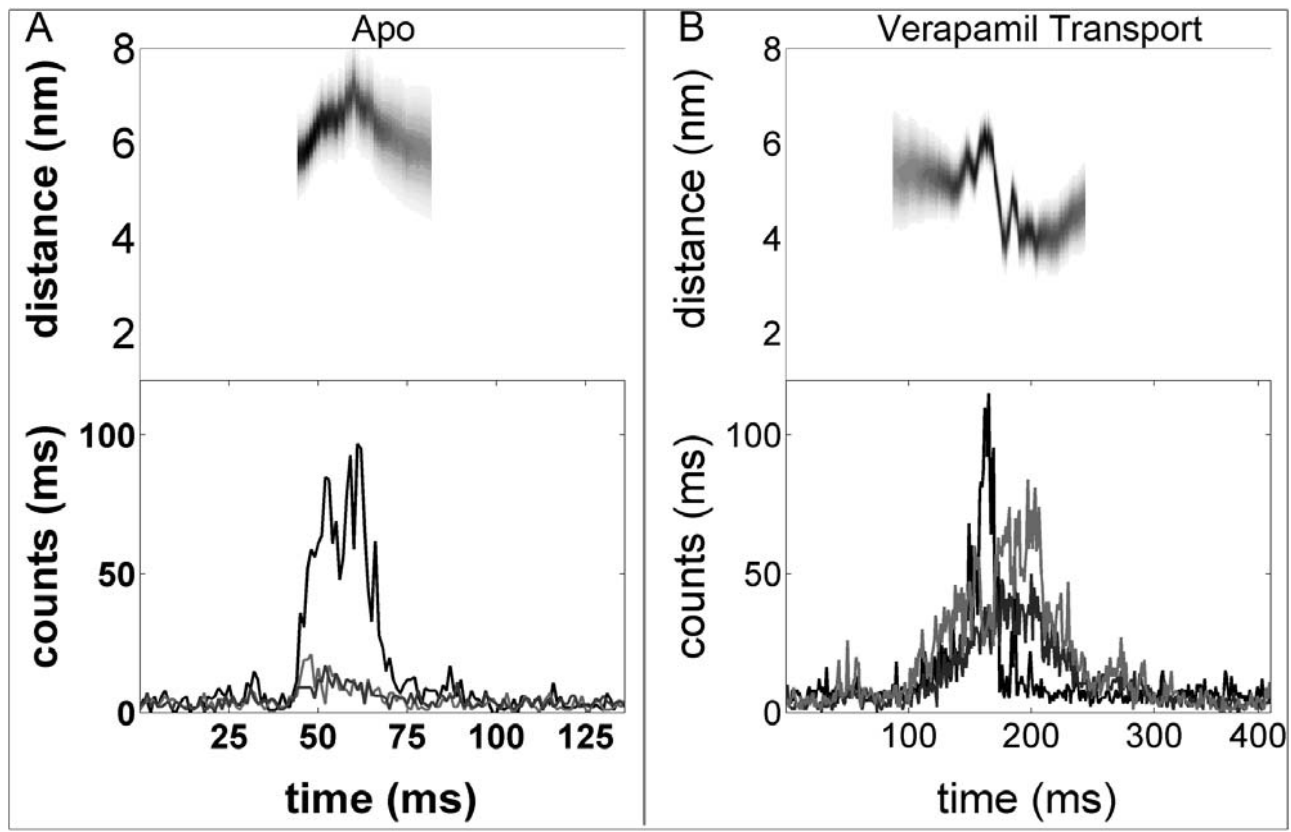

**Figure 4: Time trajectories of the FRET efficiency changes of a single Pgp transporter labeled with Alexa488 (donor) and Atto610 (acceptor) reconstituted in a liposome.** The intensity traces are background corrected, and the quantum yields and the detection efficiencies were taken into account. The black traces show the donor intensity, the light gray the FRET acceptor intensity and the dark gray traces are the acceptor test signal. (**A**) Measurement without any nucleotide or drug binding ('Apo' condition). The corresponding FRET distance trace is shown in the upper panel. The mean FRET distance of about 6.1 nm does not change with the exception of small fluctuations. (**B**) A photon burst of Pgp in the presence of 200 μM verapamil and 1mM MgATP ('Transport' condition) displays several distance changes, including one rapid step from ~6 nm to ~4 nm.

An automatic search algorithm based on intensity thresholds was applied to find the photon bursts. For the recorded data only those bursts were analyzed that met the following conditions: the length of a photon burst was between 20 and 250 ms; the acceptor test signal had to be stronger than 10 counts/ms; and the donor signal stronger than 20 counts/ms. With these search criteria we found 145 bursts for the 'Apo' condition, 908 for 'Transport', 774 for the 'vanadate trapped' and 574 for the 'cyclosporin A' conditions (vanadate trapped and cyclosporin A data not shown).

The so-called proximity factor P (see histograms in Fig. 5) correlates the intensities of the donor and acceptor channel:

$$P = \frac{I_A}{I_D + I_A}$$

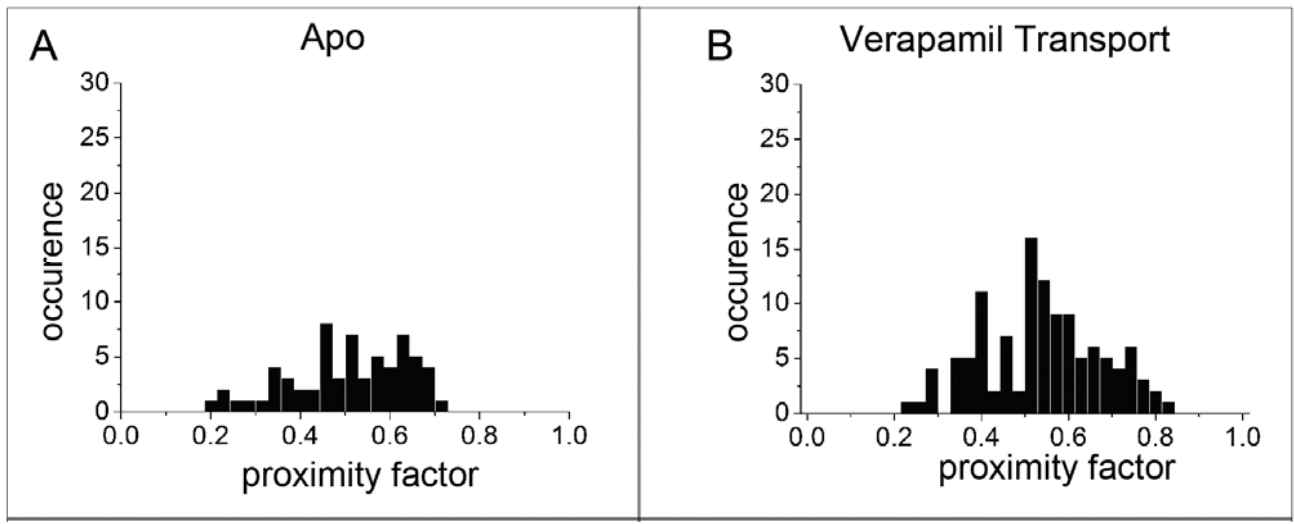

**Figure 5: Histogram of the proximity factors of Pgp transporters with two different biochemical conditions.** (**A**) Apo, no nucleotide or drug. (**B**) Transport with 200 µM verapamil and 1 mM MgATP.

As can be seen in Fig. 5, for both conditions broad distributions of FRET efficiencies were found by comparing the proximity factor histograms. In the 'Apo' condition, only a few bursts met the threshold requirements. Most of the bursts showed no or only one change in the distance between the two fluorophores. One possibility is that a larger population occupied an open conformation with a very low FRET state and was trapped in this state without any substrate binding. Proximity factors for 'Transport' condition (verapamil and MgATP) were centered around 0.5, but the histogram was broadened.

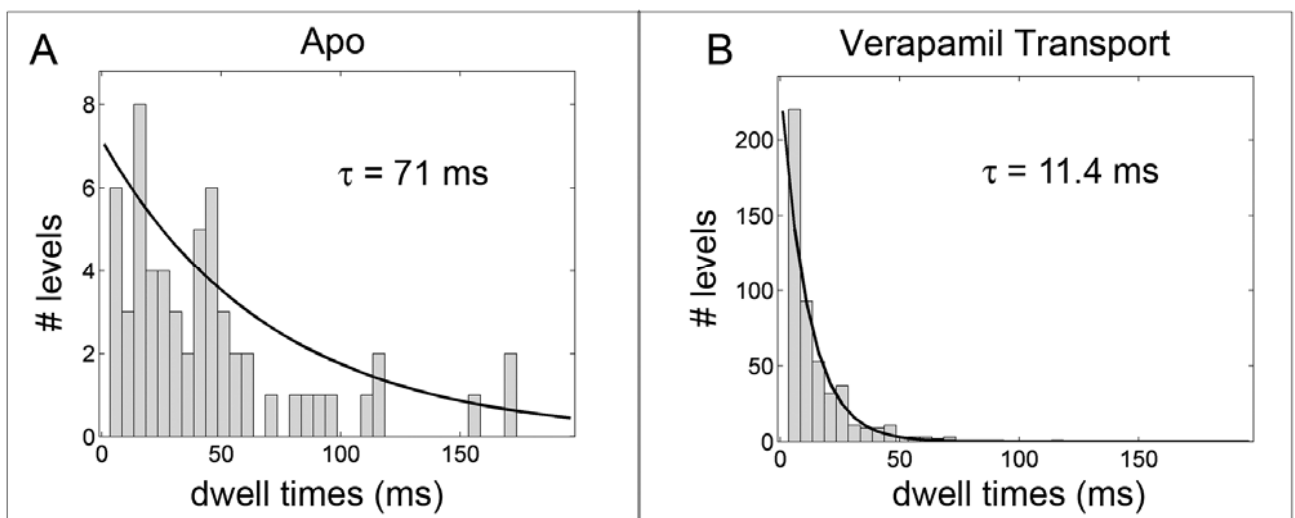

**Figure 6: Dwell time histograms for 'Apo' (A) and 'Transport' conditions (B).** The dwell time for each condition was determined by mono-exponential fitting (black curve). The dwell time binning of the histograms was 5 ms.

Fig. 6 shows the dwell time histograms of the two conditions 'Apo' and 'Transport'. The dwell time describes the time interval in which the Pgp transport cycle remains in the same conformation. From these plots, an idea about the rate of transitions that occur during different substrate conditions can be made; thus, only completely recorded steps were taken into account. Using mono-exponential decay fitting (black lines in Fig. 6), the dwell time of each condition was determined.

The 'Apo' condition appears to populate slow and quick transitions indicating a very flexible molecule, but it has to be taken into consideration that only a small amount of the recorded photon bursts showed more than one step. As expected, most of the molecules in absence of substrate occupied predominantly one conformation state. In the presence of verapamil and MgATP, the dwell times were very short indicating rapid transitions, consistent with rapid MgATP hydrolysis. Both vanadate trapping and cyclosporin slowed the molecule dynamics down, which supports previous reports of slowed ATPase transitions and limited NBD-NBD interactions, respectively[40].

In the last step of analysis, distributions of the distance changes were considered under the two conditions 'Apo' and verapamil/MgATP-driven 'Transport' (Fig. 7). The distance of the two nucleotide binding domains before (FRET distance 1) and after (FRET distance 2) a conformational change were plotted together resulting in one data point for two distances. In comparing the transition plots, it was observed that there was a broad distribution in the 'Apo' condition. However, this was based on a few FRET transitions.

In the presence of MgATP/verapamil two main transitions were found: from 5.4 nm to 5.0 nm and the reverse, as well as from 5.5 nm to 6.0 nm and back. The spots across the diagonal indicated a fluctuation between two states. This fluctuation was also present in vanadate trapped and cyclosporin conditions; however, there was only one fluctuation in each condition. The vanadate trapped molecule occupied the transitions at shorter distances whereas the cyclosporin A trapped Pgp occupied the lower FRET states (data not shown).

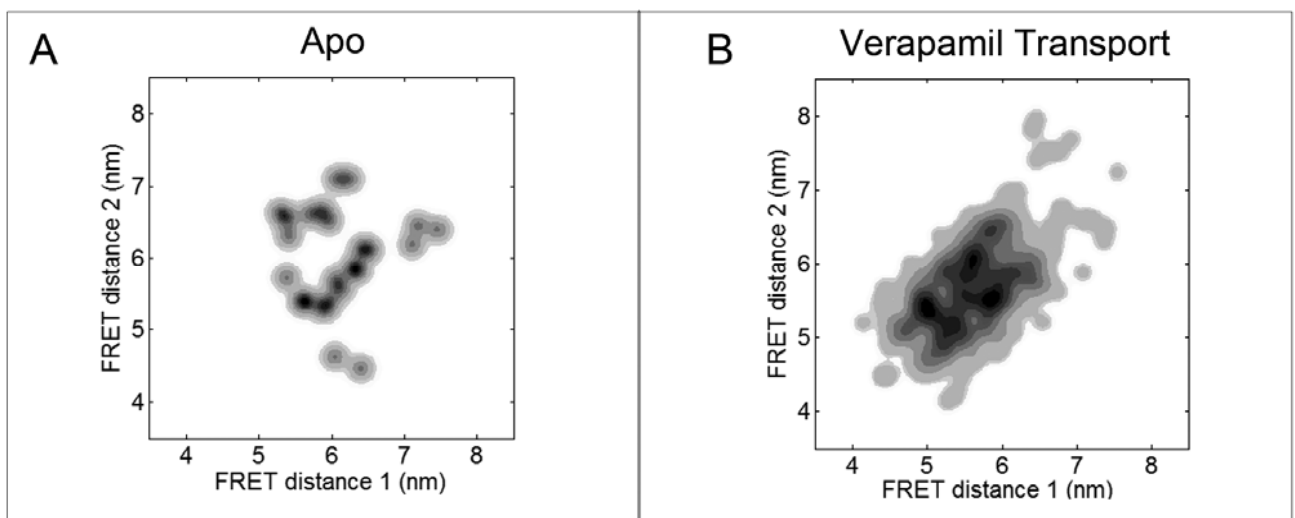

**Figure 7: FRET transition density plots of single Pgp transporters.** (**A**) in the absence of ATP and (**B**) during ATP hydrolysis. At least two distinct FRET levels had to be detected within each photon burst to be added to the histograms. In the absence of ATP only few fluctuating transporters were observed.

## 4 DISCUSSION

In order to address the questions surrounding the conformational changes during the catalytic cycle of Pgp, cysteines were placed in a cysteine-less mouse Pgp for fluorescent labeling to carry out single-molecule FRET spectroscopy. Here we show that Pgp can be labeled with two different fluorophores, reconstituted into liposomes, and FRET efficiency differences can be visualized dependent upon substrate condition. In our preliminary analysis of the double mutant TS, we observe two major conformations of the transporter, a high FRET state and a low FRET state. In addition, we see quick fluctuations in presence of verapamil and ATP in contrast to the longer dwell times of vanadate-trapped Pgp and cyclosporin activated Pgp. One limitation of the single-molecule approach used is that the observation time window (20 to 250 ms) is short compared to the average turnover rate of Pgp of ~1/s under these conditions. This might explain why only few larger distance steps were measured.

To overcome these observation time limitations of the combination of freely diffusing Pgp proteoliposomes and the relatively slow turnover by Pgp, a trapping technique has to be applied. A promising new approach is the ABELtrap which uses electrokinetic forces to hold a liposome in solution. The position of a (fluorescently labeled) proteoliposome is monitored by an EMCCD camera to generate electrical potentials at electrodes in order to compensate the Brownian motion[41, 42]. Additionally, Hidden Markov Models (HMM) analysis[43] will provide a more objective way of analyzing the FRET time trajectories of Pgp to link together the transitions of NBDs of Pgp, and will allow for the next steps in furthering the understanding of the step-by-step mechanism of Pgp.

**Acknowledgements**

This work was in part supported by NIH grants CA100246 and GM058600 to S.W.